\documentclass[sigconf,10pt]{acmart}

\usepackage{graphicx,subfigure}
\usepackage{amsmath,amsthm,amssymb}
\usepackage{units}
\usepackage{algorithm,algorithmic}

\usepackage{xcolor}
\usepackage{gensymb}
\usepackage{nicefrac}
\usepackage{hyperref}

\usepackage[utf8]{inputenc}
\usepackage[T1]{fontenc}

\setcopyright{none}

\acmISBN{none}
\acmDOI{none}

\title{Dead on Arrival: An Empirical Study of The Bluetooth~5.1 Positioning System}

\author{Marco Cominelli}
\affiliation{%
	\institution{University of Brescia}
	\state{Italy}
}
\email{marco.cominelli@unibs.it}

\author{Paul Patras}
\affiliation{%
	\institution{The University of Edinburgh} 
	\state{Scotland, UK}
}
\email{ppatras@inf.ed.ac.uk}

\author{Francesco Gringoli}
\affiliation{%
	\institution{CNIT/University of Brescia}
	\state{Italy}
}
\email{francesco.gringoli@unibs.it}

\begin{document}


\setcopyright{none}

\acmISBN{none}
\acmDOI{none}

\settopmatter{printacmref=false}

\fancyhead{}

\begin{abstract} 
The recently released Bluetooth 5.1 specification introduces fine-grained positioning capabilities in this wireless technology, which is deemed essential to context-/location-based Internet of Things (IoT) applications. In this paper, we evaluate experimentally, for the first time, the accuracy of a positioning system based on the Angle of Arrival (AoA) mechanism adopted by the Bluetooth standard.
We first scrutinize the fidelity of angular detection and then assess the feasibility of using angle information from multiple fixed receivers to determine the position of a device. Our results reveal that angular detection is limited to a restricted range. On the other hand, even in a simple deployment with only two antennas per receiver, the AoA-based positioning technique can achieve sub-meter accuracy; yet attaining localization within a few centimeters remains a difficult endeavor. We then demonstrate that a malicious device may be able to easily alter the truthfulness of the measured AoA, by tampering with the packet structure.  To counter this protocol weakness, we propose simple remedies that are missing in the standard, but which can be adopted with little effort by manufacturers, to secure the Bluetooth 5.1 positioning system.
\end{abstract}

\maketitle

\section{Introduction}
Indoor localization of user devices is a critical research topic that has been attracting increasing interest from vendors~\cite{meraki}, app developers~\cite{ibeacon}, and the researcher community at large~\cite{zafari:2019}. Localization is also a key application on the 5G mobile technology roadmap~\cite{witrisal:2016}.
To date, however, a positioning solution as widespread as GPS and usable when satellite signals cannot be received has not been available. 

To address this problem in IoT scenarios, the Bluetooth Special Interest Group (Bluetooth SIG) introduced a set of features in the latest Bluetooth Core Specification v5.1~\cite{BT5.1}, which are specifically aimed at determining the location of a device with high accuracy. In particular, the standard adopts two signal processing techniques for identifying the Angle-of-Arrival (AoA) and  Angle-of-Departure (AoD) of a transmitted signal. AoA enables a receiver to determine the angular position of a transmitter by measuring the phase-delay at multiple antennas. With AoD, a transmitter having multiple antennas can send a signal that allows receivers equipped with a single antenna to detect their angular position with respect to the transmitter. Combined with distance estimation~\cite{chowdury:2015,bertuletti:2016}, these techniques aim to help pinpoint the precise location of a device.

Numerous systems based on different technologies have been proposed to date to tackle indoor localization, ranging from those whereby users carry smart tags, to systems that opportunistically use signals transmitted by mobile devices/smartphones to infer their position. Naturally, previous solutions build upon wireless communications systems commonly embedded into mobile devices, including Wi-Fi~\cite{luo:2017,vasisht:2016}, Bluetooth~\cite{ibeacon_rssi,ibeacon_pos_hospital}, and ultra-wideband (UWB)~\cite{kempke2015polypoint,kempke2016surepoint} transceivers. Nevertheless, even if some of these solutions demonstrated remarkable performance in terms of positioning accuracy, none of them gained enough traction to witness wide adoption. With the growing adoption of IoT technology and the emergence of standardized methods for positioning, the situation is bound to change. In 2018, nearly 4 billion devices were shipped with Bluetooth technology and, thanks to its low energy capability, the Bluetooth SIG forecasts that the location services domain will encompass over 400 million products per year by 2022~\cite{BTSIG-press}, with applications spanning supply chain asset tracking, customized visitor experience in museums through proximity detection, smart homes, healthcare, and many more.

\textbf{Contributions.} In this paper we test the market readiness of the Bluetooth 5.1 positioning capability by experimentally evaluating the performance of the adopted AoA mechanism from two perspectives: that of pure angular measurement accuracy and the ability to correlate two or more angular measurements in order to estimate a device's position in a 2D plane. To this end, we use a software-defined radio (SDR) testbed and deploy the BLE 5.1 positioning technique in its simplest form, i.e., with only two antennas at the receiver.\footnote{At the time of writing, commercial devices implementing Bluetooth 5.1 positioning were not launched to market. We thus resort to SDR hardware.} We report results that offer a first glimpse into the performance of this localization solution, and a primer for more complex implementations that are yet to appear.

Through our study, we first reveal severe limitations that affect angular measurements and which restrict the applicability of the AoA technique within a specific circular sector centered at the receiver. Secondly, we show that positioning based on AoA measurements, although offering sub-meter accuracy, is far from achieving centimetre-level precision. Our findings should prove useful to system and app developers who aim to build upon this feature. We then provide a preliminary assessment of the (in)security of the AoA-based positioning mechanism, laying out guidelines on antenna switching patterns that manufacturers could follow to prevent attackers from compromising position truthfulness.
Finally, we release the tool opensource, interested readers can download and test it from \href{https://github.com/bsnet/bleaoa}{https://github.com/bsnet/bleaoa}.

\section{Related work} \label{sec:rework}

Determining a wireless device's position should be strictly a matter of signal direction (angle) finding. The problem of estimating the Angle-of-Arrival (AoA) of a signal has been extensively studied. In general, an antenna array is required in order to measure the phase-delay between the replicas of the signal received by each element of the array. The most common approach to determine the AoA based on the measured phase-delay is the multiple signal classification (MUSIC)~\cite{schmidt_music}, which achieves excellent angular resolution.

In commodity wireless systems, however, estimating the position of a transmitter has been largely based on measuring the power of the received packets (RSSI). With respect to Bluetooth Low Energy (BLE), the accuracy of positioning frameworks based on iBeacon technology has been studied in \cite{ibeacon_rssi} and \cite{ibeacon_pos_hospital}. In the former, an average localization error of 4~m was achieved by installing 36 beacons. The latter divided a testbed into 12 subareas and obtained localization errors within 5~m of adjacent subareas. An analysis of how positioning accuracy depends on the number of BLE beacons has been carried out in \cite{ji:2015}.
More recently, De Blasio et al. examined the positioning accuracy of BLE~5.0 in a deployment with 12 beacons in a 168~$\mathrm{m}^2$ area, reporting accuracy within 2.5~m~\cite{deblasio:2018}. 
The key limitation of existing methods for estimating indoor position is that they assume an accurate channel model, which is very difficult to build. Moreover, different BLE channels may exhibit different characteristics, leading to modest positioning accuracy when relying on RSSI~\cite{powar:2017}. To circumvent these problems, MUSIC has been applied recently to determine the position of BLE transmitters based on the AoA estimated by multiple nodes~\cite{ble_music_positioning}.

The decision made by the Bluetooth SIG to include the direction finding feature in the new BLE standard reshapes the positioning problem. In particular, in order to apply MUSIC, multiple coherent RF channels would be required. Instead, the AoA feature in BLE 5.1 uses only one channel connected to the different elements of an antenna array and an RF switch to select among them~\cite{BT5.1}. The sequence transmitted to perform AoA measurements is assumed to be known (as described in the next sections). Luo et al. have conducted simulations to assess the performance of AoA estimation in the case of known transmit sequences~\cite{simulation_aoa}. However, a thorough characterization of the positioning accuracy of BLE~5.1 in real deployments has not been reported yet. This is mainly due to the lack of commercial hardware supporting this feature. 
To the best of our knowledge, our work is the first to conduct an experimental study of the BLE 5.1 positioning system and document its performance and vulnerabilities.

\section{Bluetooth Low Energy (BLE)}

BLE is a wireless standard designed for inexpensive personal area networks that require low power consumption and low data rates. 
BLE has been integrated into version 4.0 of the Bluetooth Core specification in 2010 and has been also marketed as Bluetooth Smart.

At the physical layer (PHY), BLE operates in the 2.4 GHz ISM Band. The access to the medium is regulated by a hybrid time-frequency division multiplexing scheme. In particular, the assigned 80 MHz bandwidth is divided into 40 orthogonal channels with central frequencies equally spaced by 2~MHz. Two types of BLE channels exist: 3 \textit{advertising} channels are used for enabling device discovery, connection setup, and broadcasting messages; the other 37 \textit{data} channels are used to exchange data. 
When a connection is established between a pair of devices, an Adaptive Frequency Hopping (AFH) technique is used to combat interference: connected devices switch rapidly between channels according to a pseudo-random sequence that is known by both transmitter and receiver.
Channels can be dynamically removed from the hopping sequence depending on external conditions (e.g., strong narrow-band interference). The communication between devices happens at specific time intervals. A channel access policy is defined based on time slots and intervals that depend on the role of a BLE device (master/slave).

For transmission, BLE employs binary Gaussian Frequency Shift Keying (GFSK) modulation with a bandwidth-bit period product of 0.5 and two possible symbol rates: 1 Msym/s and 2 Msym/s. Even if four different PHY modes build on these modulation schemes, in version 5.1 of the BLE standard the AoA mechanism can be used only with the Uncoded PHYs. In the rest of the paper we confine consideration to the mandatory LE~1M~PHY with 1~Mb/s data rate.

\begin{figure}[t]
\centering
\includegraphics[width=0.85\columnwidth]{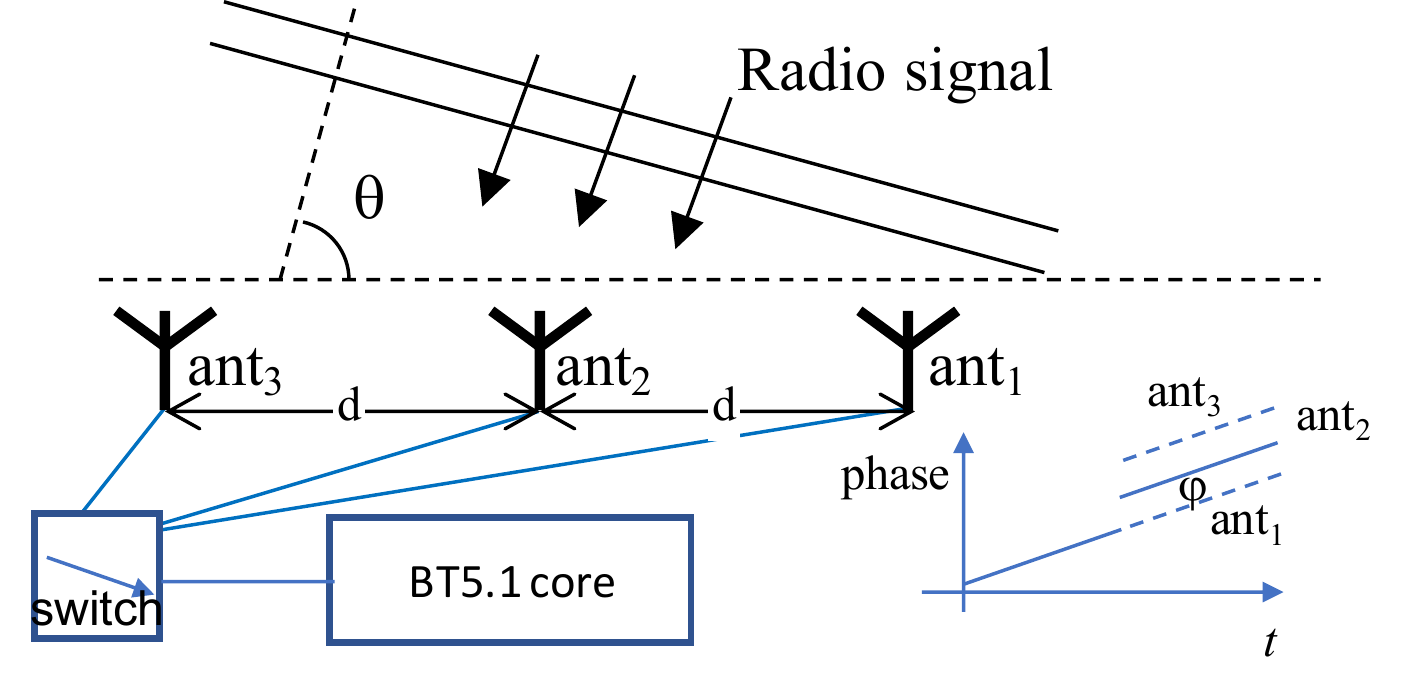}
\caption{The AoA mechanism needs to {\bf estimate} phase-delay $\varphi$ between antennas to compute angle~$\theta$: multiple antennas are connected to a single radio transceiver using an RF switch.}
\label{fig:aoa}
\end{figure}

\subsection{Direction Finding in BLE 5.1}
According to the standard, a device equipped with an antenna array of $M$ elements can determine the AoA of signals from a transmitter using simple geometric calculations. The documentation considers the scenario illustrated in Fig.~\ref{fig:aoa}. Assuming the radio signal is a plane wave with constant frequency impinging on the antenna array of the receiver, the phase difference $\varphi$ between the signals received at each pair of adjacent antennas is expressed as 
\begin{equation}
\varphi = 2 \pi (d / \lambda) \cos \theta,
\label{eq:phase}
\end{equation}
where $\lambda$ is the wavelength of the signal, $d$ the distance between the antennas, and $\theta$ the angle of arrival. Therefore,
\begin{equation}
\theta = \arccos \left( \frac{\lambda \varphi}{2 \pi d} \right) \, .
\label{eq:aoa}
\end{equation}
By measuring the phase-delay $\varphi$ and knowing both $\lambda$ and $d$, computing $\theta$ is straightforward.
Fusing $\theta$ values computed at different antenna pairs is left to the manufacturer.

From Eq.~\ref{eq:phase} it is clear that all the angles $\theta$ from $0^\circ$ to $180^\circ$ can be determined from the phase-delay $\varphi$ only if $d < \lambda/2$. If this condition is not met, then an aliasing phenomenon appears, whereby it is not possible to uniquely map a value of $\varphi$ to an angle $\theta$.
Further, how to evaluate the phase-delay $\varphi$ is unclear, because the standard requires a single radio in the chipset to be connected to the different antennas using a RF switch (as in Fig.~\ref{fig:aoa}) and a procedure for inferring the phase-delay is not specified in the official documentation. Instead,  manufacturers can develop their own algorithms to estimate $\varphi$. We explain in the next section how we implemented this feature on our SDR platform. On the other hand, the standard specifies \emph{(i)} the format of the field inside a packet that should be used to evaluate the phase-delay and \emph{(ii)} the timing for performing antenna switching over this field. We discuss these features next.


\subsection{Direction Finding Packets and Antenna Switching Timings}

\begin{figure}[h]
\centering
\includegraphics[width=0.97\columnwidth]{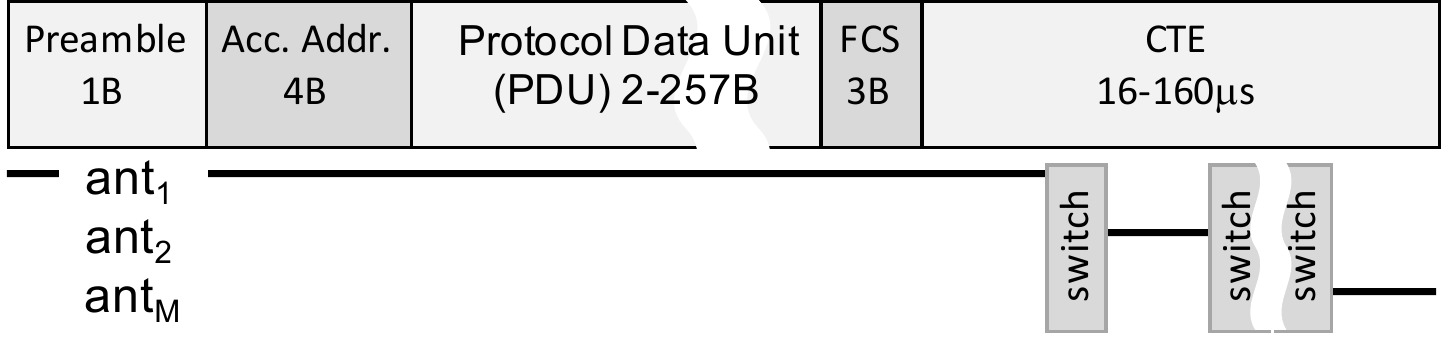}
\caption{The format of a packet supporting AoA detection, and corresponding switching timing.}
\label{fig:packet}
\end{figure}

BLE packets supporting the direction finding capability embed an additional field called Constant Tone Extension (CTE) that follows the CRC coefficient as we show in Fig.~\ref{fig:packet}.
The CTE consists of a constantly modulated sequence of unwhitened {\sc 1}-valued bits, i.e., a constant tone signal with variable length, which can last between 16--160~$\mathrm{\mu s}$. This is divided into different subfields: a {\bf reference period} ($8~\mathrm{\mu s}$) is sent first, after a guard interval; then, an alternating sequence of {\bf switch slots} and {\bf sample slots} follows. Slots of $2~\mathrm{\mu s}$ must be implemented by all the devices that support the direction finding features, whereas slots of $1~\mathrm{\mu s}$ can be optionally supported.

A receiver uses one antenna to receive a BLE packet (from preamble to the CRC field) and relies on the same antenna to collect 8~IQ samples during the reference period ($8~\mathrm{\mu s}$ sampled at 1~MS/s). It then switches among the available antennas during switch slots, taking one IQ sample per sample slot (even if the sample slot is $2~\mathrm{\mu s}$-long). The switching pattern is defined by the BLE host. The shortest possible switching pattern lasts $16~\mathrm{\mu s}$ and uses only two antennas. This is also the pattern with which we work in our study. We leave investigating the impact of the length of the switching pattern on positioning accuracy for future work.

\section{BLE 5.1 Testbed Deployment}

We implement Software-Defined Radio (SDR) prototypes to replicate the behavior of BLE transceivers supporting the AoA detection mechanism as defined by the Bluetooth 5.1 Specification.
Our setup consists of two USRP Ettus B210 boards that we use for receiving, and a USRP Ettus N200 that we use for transmitting. The receiver is connected to a laptop powered by an Intel i5 CPU clocked at 2.7GHz with 8GB of RAM, which has enough power to run our software receiver in real-time.
We manufactured a plastic support to place two half-wavelength dipole antennas at a distance of 6~cm from each other; each antenna is connected to a TX/RX port of the USRP B210 using a rigid coaxial cable.
The transmitter is driven by a Chromebook powered by an Intel Core~M CPU clocked at 2GHz with 2 GB of RAM. All computers run Ubuntu 18.04. Next, we describe our implementation of the BLE AoA detection mechanism, introducing first the real-time BLE software transmitter/receiver developed, then explaining how we customize this to emulate the AoA detection feature.

\subsection{Emulating BLE 5.1 Connections}

For our experiments we do not setup a real BLE FH data connection; we rather emulate it by continuously transmitting packets and tuning all nodes simultaneously on the same channels. We program the software transmitter to generate BLE packets, which we encode as LE~1M~PHY and send at the rate of 100 packets per second, using a fixed Access Address. Inside the payload we embed a sequence number that we use for debugging purposes and for matching the same packet at multiple receivers. To emulate the CTE, we add a sequence of binary ones at the end of each packet.

The software receiver acquires IQ samples with a sampling frequency of 2~MS/s. It then decodes bits at 1~Mb/s by operating a phase-discrimination procedure on consecutive pairs of samples. Finally, it detects valid packets starting from every preamble found and checking the validity of the CRC. 

It is important to notice that in BLE~5.1 the CTE is not subject to error checking and that in the packet format (Fig.~\ref{fig:packet}) it comes right after the CRC field. In addition, AoA measurements can be performed by BLE devices even if errors occurred while receiving the packet.

To achieve FH and ensure the transmitter and receivers are on the same physical channel, we rely on out-of-band signalling performed via a wired network that connects all nodes and distributes information generated by a controller. The controller provides to all nodes a deterministic hopping sequence spanning all the available BLE channels.
Each receiver can adjust the gain dynamically by measuring the amplitude of the IQ samples in the received BLE packets, so that it can use the entire dynamic range of the Analog-to-Digital Converters of the USRP B210.

\subsection{Implementing AoA Detection} \label{sec:aoa}

As described before, BLE devices supporting the AoA mechanism have only one receiving radio chain that is connected to the different elements of the antenna array using an RF switch. This means that only one antenna can be active at any given time and that phase-delay $\varphi$ must be inferred from IQ samples captured during the reference period in the CTE and in the following sample slot. We explain here the algorithm that we implemented for inferring the phase-delay and how we emulated it over USRP B210 boards. Hereafter we consider the case with $M=2$ antennas.

\begin{figure}[h]
    \centering
    \includegraphics[width=0.90\columnwidth]{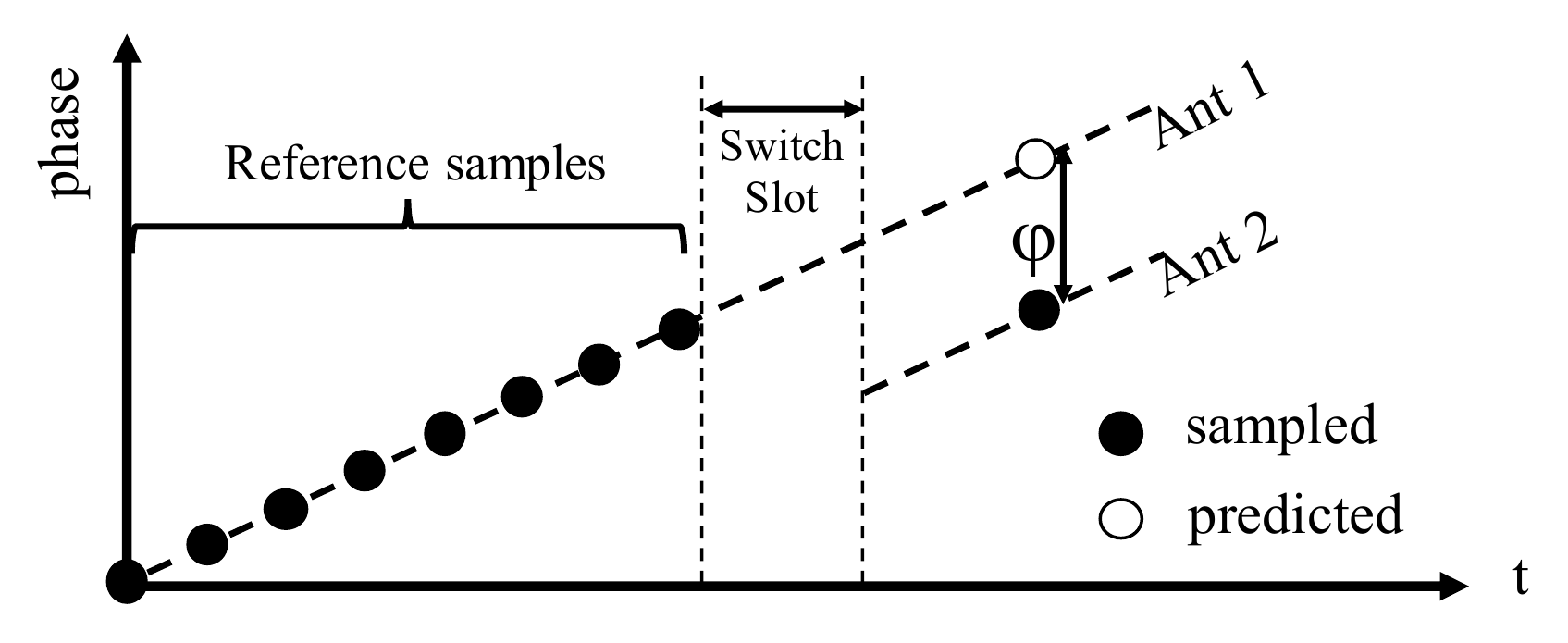}
    \caption{The phase difference is computed by subtracting the phase sampled on antenna 2 from the phase predicted on antenna 1 according to the samples it received during the reference period.}
    \label{fig:phase_evo}
\end{figure}

As the CTE is a sequence of unwhitened ones, the signal appears as a constant frequency tone and as such its phase increases linearly. During the CTE reference period, we collect 8 IQ samples from antenna 1 and we build a linear model of its phase evolution. The AoA antenna switching takes place during the switch slot, after which we collect an IQ sample from antenna 2 exactly at the next sample slot. We then compare the phase of this IQ sample with the instantaneous phase of the signal on antenna 1 that we estimate using the linear model built during the reference period, as shown in Fig.~\ref{fig:phase_evo}. The phase difference estimated with this method corresponds to $\varphi$ in Eq.~\ref{eq:phase}, and the angle of arrival $\theta$ is found with Eq.~\ref{eq:aoa}.

Emulating the algorithm over the B210 boards is straightforward: we continuously look for valid  packets received at one antenna and, once we detect one, we process the reference period of the CTE, we predict the value of the phase on this antenna in the following sample slot, and we finally subtract from it the value of the phase sampled on the other antenna at that same moment in time.


In our setup, we account for a constant phase offset between the two receiving chains of each B210 board. This delay can result from path differences between the two receiving chains due to the specific manufacturing of the boards and the antennas used. Before deploying the boards in the testbed, we execute a calibration procedure by connecting both input ports of a USRP to the same source with a splitter, using good quality cables of the same length.

\subsection{Positioning Using BLE 5.1}\label{sec:positioning}

\begin{figure}
\centering
\includegraphics[width=0.75\columnwidth]{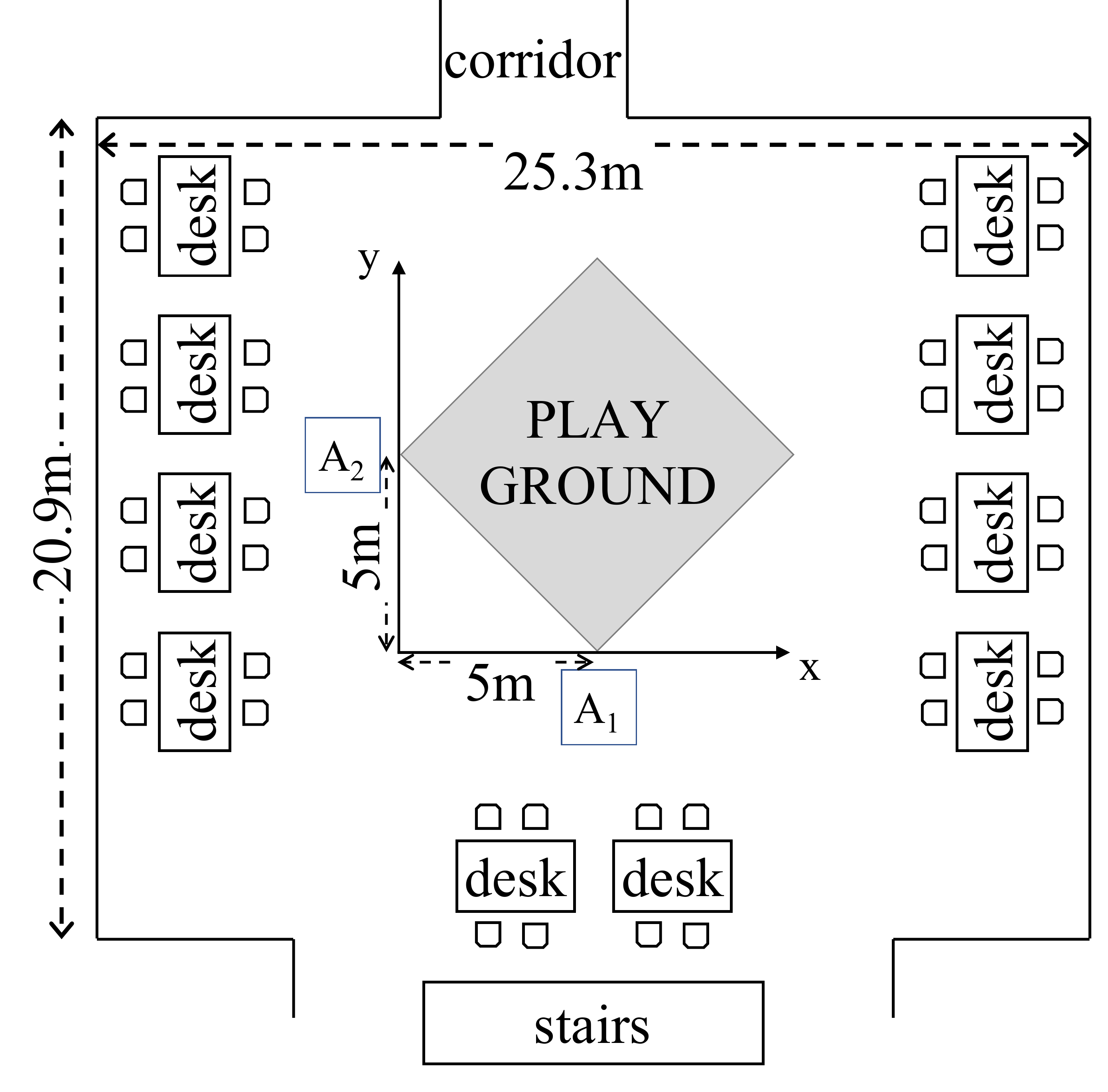}
\caption{Indoor testbed used for evaluating the AoA-based positioning mechanism. Two receivers (anchors) $A_1$ and $A_2$ determine the position of target transmitter in the ``playground''.}
\label{fig:playground}
\end{figure}

To estimate the position of the transmitter, instead of ranging, we use an additional USRP B210 connected to a different laptop with same specification as before.
The two software receivers perform the steps described before to determine AoAs independently. The receivers are connected in a network, so that it is possible to collect all the AoA measurements and use simple geometric operations to convert the two angles into the 2D coordinates of the target device. 
The configuration we used for positioning experiments is the indoor scenario shown in Fig.~\ref{fig:playground}.
The two receivers (\textit{anchors}) $A_1$ and $A_2$ are placed at equal distance from the origin of the reference frame. The linear antenna array of each receiver is aligned with the $x$ and $y$ axes of the reference frame respectively.

\section{Experimental Results}
We test the viability of the positioning mechanism adopted by BLE 5.1 by conducting two sets of experiments: one focusing on assessing the accuracy of the determined AoAs,~\mbox{the other} on quantifying positioning errors when employing 2~\mbox{anchors.}

\subsection{Angular Accuracy}
We begin by evaluating the angular accuracy in the simplest configuration, i.e., a device running our software receiver with two dipole antennas, and another running our software transmitter. To isolate the impact of reflections and multipath-induced errors, we ran these experiments in an outdoor scenario, on a flat court far away from obstacles, with both the receiver and transmitter placed at 0.5 m above the ground and all antennas of the same type, i.e., dipoles positioned vertically.


We measure the angle $\theta$  between the axis of the antenna array and the propagation direction of the signal, starting with $\theta = 90\degree$, which corresponds to ideal no-phase-delay. We progressively reduce the angle to $0\degree$, in $5\degree$ steps. For all angles we collect 30 correctly received packets with CTE extension, on each of the 40 BLE channels, which corresponds to a total of 1,200 phase-delay measurements per angle. As the two antennas are spaced less than half of the wavelength for all the BLE channels, and given the position of the reference antenna, by reducing $\theta$ we expect to observe the phase-delay decreasing monotonically. 

\begin{figure}
\centering
\includegraphics[width=0.88\columnwidth]{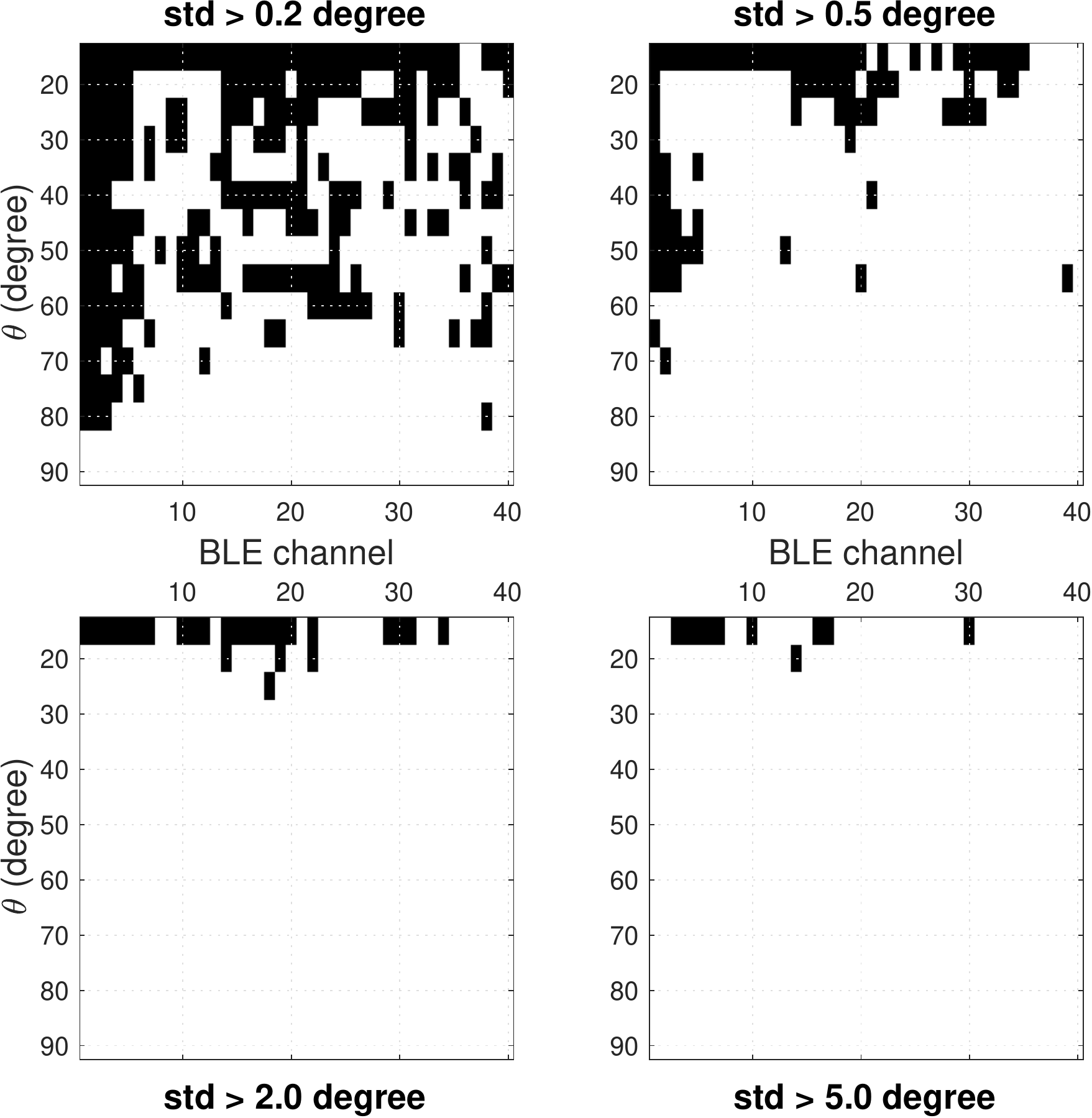}
\caption{Channel (x-axis) and angle (y-axis) combinations for which the standard deviation of the estimated angle over 30 measurements is below (white) or exceeds (black) the standard deviation threshold indicated in the sub-plot title.}
\label{fig:angularstd}
\end{figure}

First of all, we note that, when the propagation direction of the signal is close to the axis of the antenna array, the collected phase-delays are almost random. Hence we do not report data for the range $0\degree \le \theta \le 10\degree$.
We then check the accuracy of the estimated angle within each of the $40$ BLE channels, for all the remaining 16 values, i.e.,
$\theta \in \{ 15\degree, 20\degree, ..., 90\degree \}$.
The four bitmaps in Fig.~\ref{fig:angularstd} demonstrate that the estimated angle is relatively stable over different packets received on the same channel/angle. For instance (bottom maps), only $1.5\%$ of the explored configurations exceed a standard deviation of $5\degree$, and only $4\%$ exceed $2\degree$ (this happens almost only for $\theta = 15\degree$).  More than $65\%$ of the configurations exhibit a standard deviation below $0.2\degree$ (top-left map) and it is interesting to notice that estimation over higher frequency channels seems to be more accurate.

Looking more closely at the data collected, even though the qualitative variation with $\theta$ is correct, we observe two issues: \emph{(i)} angle estimation depends on the channel, this becoming more noticeable at lower frequencies, i.e., estimations spread more; and \emph{(ii)} because of spreading, for angles close to zero there is a higher chance the phase-delay wraps around and the estimated angle bounces up to $180\degree$. 
Interestingly, for $60\degree \le \theta \le 90\degree$ the dependency with channel seems {\it random} rather than deterministic: lower frequency channels, in fact, spread similarly to higher frequency channels, suggesting there is some residual multipath effect affecting the estimation on all channels in the same way. Instead, for the $40\degree \le \theta \le 55\degree$ range, estimations spread much less. This is somewhat expected, since for such angles the size of the antenna array seen by the incident wave is much smaller than in the $\theta = 90\degree$ case and waves with higher frequency and smaller wavelength can be more accurate.
To avoid wrap-around phenomena on the phase difference that greatly affects the angle error, we exclude from the rest of the analysis situations where $\theta$ can be small: hence, we limit the ``cone'' by considering only $\theta \ge 35\degree$.
The {\it random dependency} with the frequency suggests that averaging estimations obtained on different channels would be appropriate, to reject the uncertainty due to multipath. This would come at no cost, given that AoA should be used within connection events, i.e., when the two AoA nodes are hopping over the channel sequence that was decided during the connection establishment phase. For this reason in the following we consider only data channels, excluding those dedicated to advertising.


\begin{figure}[t]
\centering
\includegraphics[width=0.97\columnwidth]{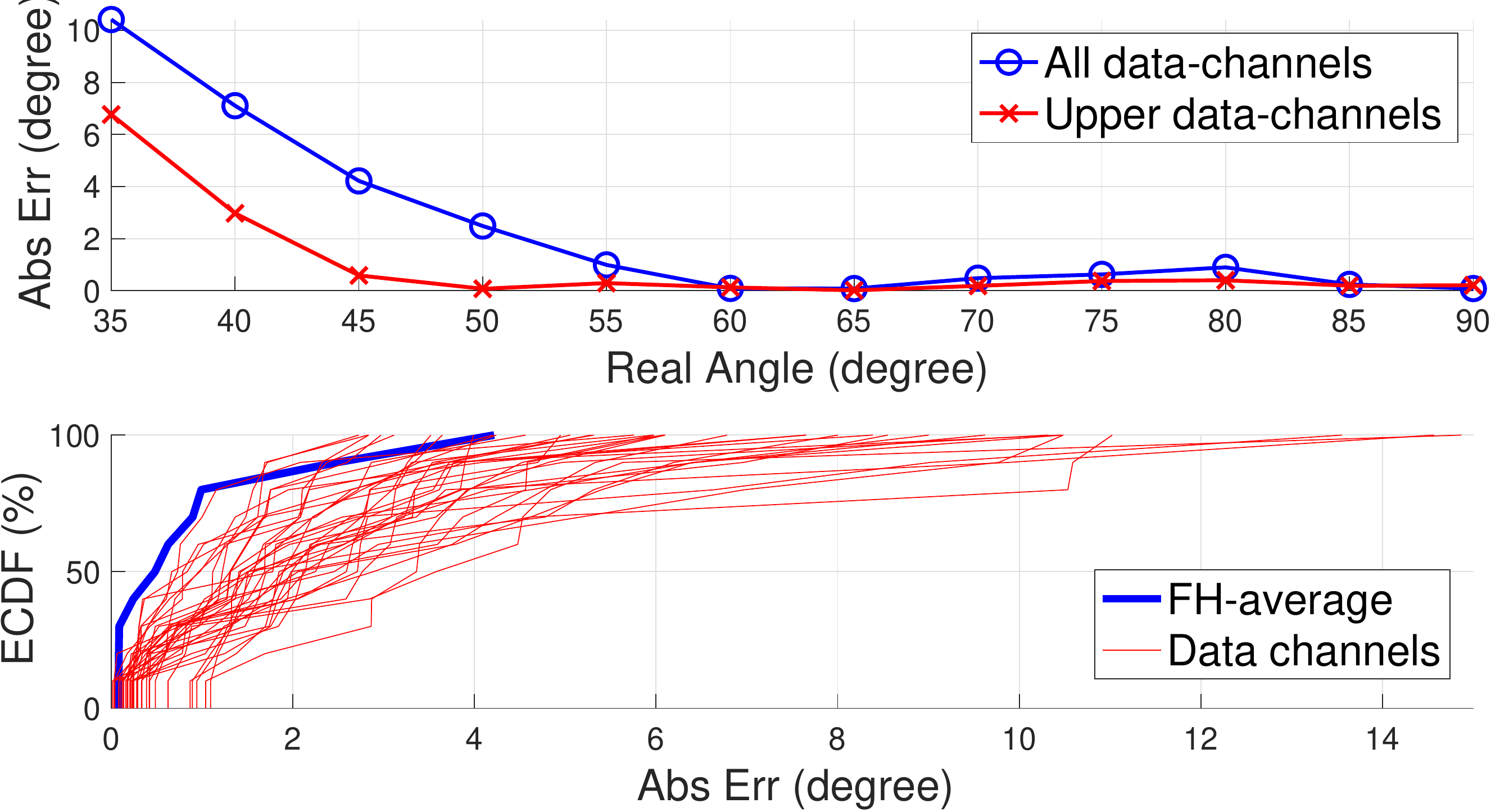}
\caption{Estimation errors in outdoor settings. Top: absolute error after averaging over all data channels (blue) and over upper-frequency half (red); bottom: ECDF of error after averaging over all data channels and for each channel; angles of $35\degree$ and $40\degree$ removed.}
\label{fig:grasserror}
\end{figure}

Fig.~\ref{fig:grasserror}-top, where we report the absolute estimation error after averaging over different sets of data channels, confirms our finding: while for $60\degree \le \theta \le 90\degree$ restricting the average over the upper half of the set of data channels does not bring any advantage, using only higher frequencies would be beneficial for higher rotations, i.e., for $\theta \le 55\degree$. However, nodes may exclude upper channels from the FH sequence and for this reason in the following we will always consider all data channels. We will instead limit further the maximum rotation by restricting even more the ``cone'', setting $\theta \ge 45\degree$, to contain the maximum error within $4\degree$. In the bottom part of the figure, we show the ECDF of the absolute error within this new cone. To this end, we consider the estimation after averaging over all data channels (thick blue line) and that of every data channel considered alone (thinner red lines). It can be noticed that apart from very few cases (some red lines above the average ECDF close to $100\%$) averaging over all data channels is always beneficial. We also note that $80\%$ of the averaged estimations are affected by error below $1\degree$.

Before moving to {\it positioning}, we repeat this experiment in the indoor scenario. This time however we keep the receiver fixed and we move the transmitter along a straight line placed at 4~m from the receiver, in 40~cm steps. 
This measurement procedure is much more similar to what we will consider next, i.e., it faces different propagation issues at different positions of the transmitter because of the stronger multipath effects inherent to indoor environments as that in Fig.~\ref{fig:playground}. As this environment is not symmetric and there are plenty of objects around, we reduce the cone to $50\degree < \theta < 130\degree$.


\begin{figure}
\centering
\includegraphics[width=0.97\columnwidth]{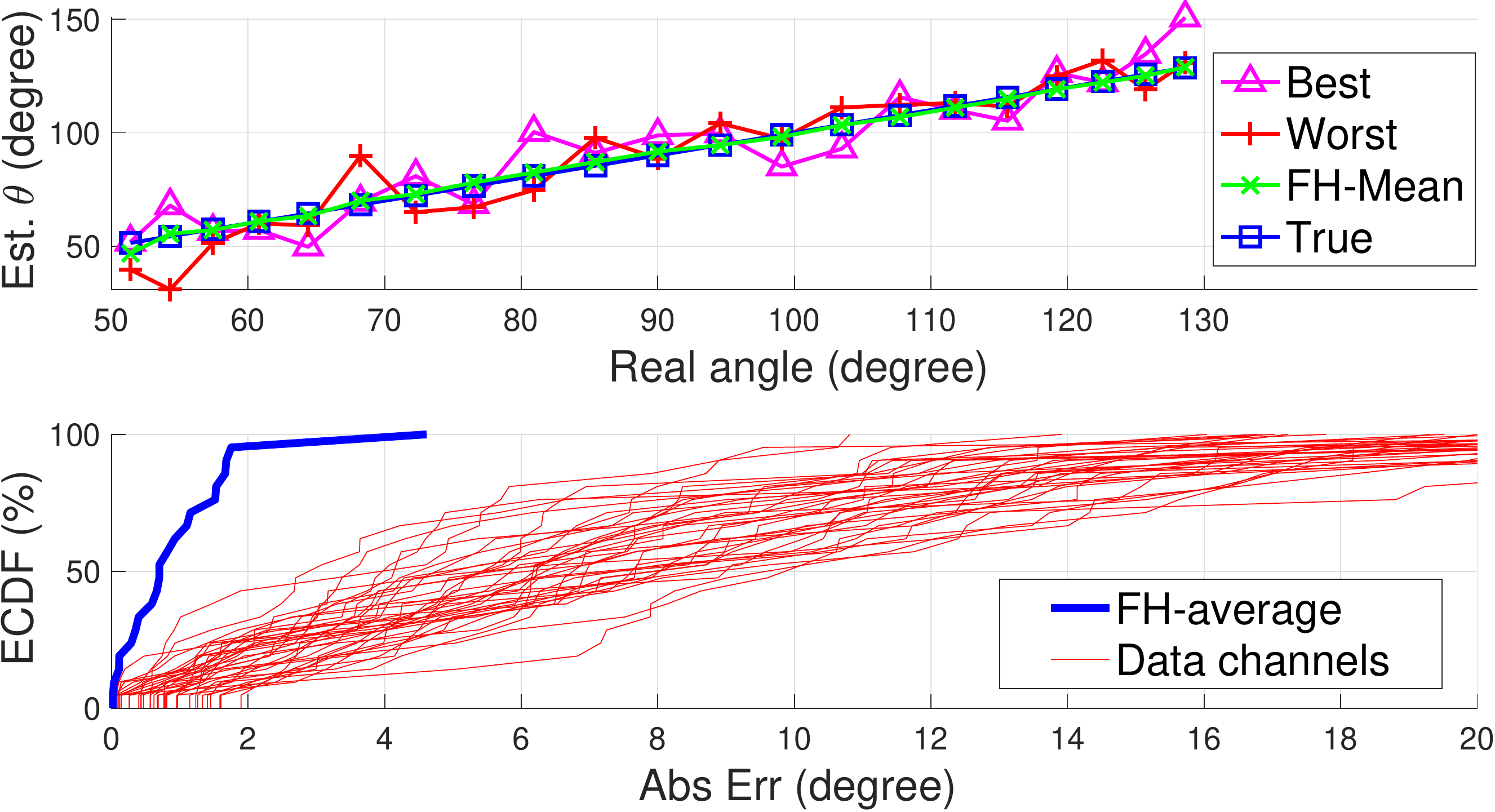}
\caption{Indoor results for angular accuracy: top, qualitative match of average estimation to {\it True} angle; bottom, ECDF of absolute error.}
\label{fig:statsindoor}
\end{figure}

We report the obtained estimation results in Fig.~\ref{fig:statsindoor}. In the top part we emphasize the very good qualitative match between the {\it True} angle and the one estimated after averaging over all data channels. We also show the estimations obtained on selected channels, respectively the one with the minimum and maximum root-mean-square error computed over all the considered angles. The benefit of averaging over channels becomes evident. In the bottom part we give a comparison of the ECDF after averaging and for every channel. Wrt. the outdoor case, we note a much worse estimation on each separate channel, which however reduces to similar results when averaging is performed.

\begin{figure}
\centering
\includegraphics[width=0.97\columnwidth]{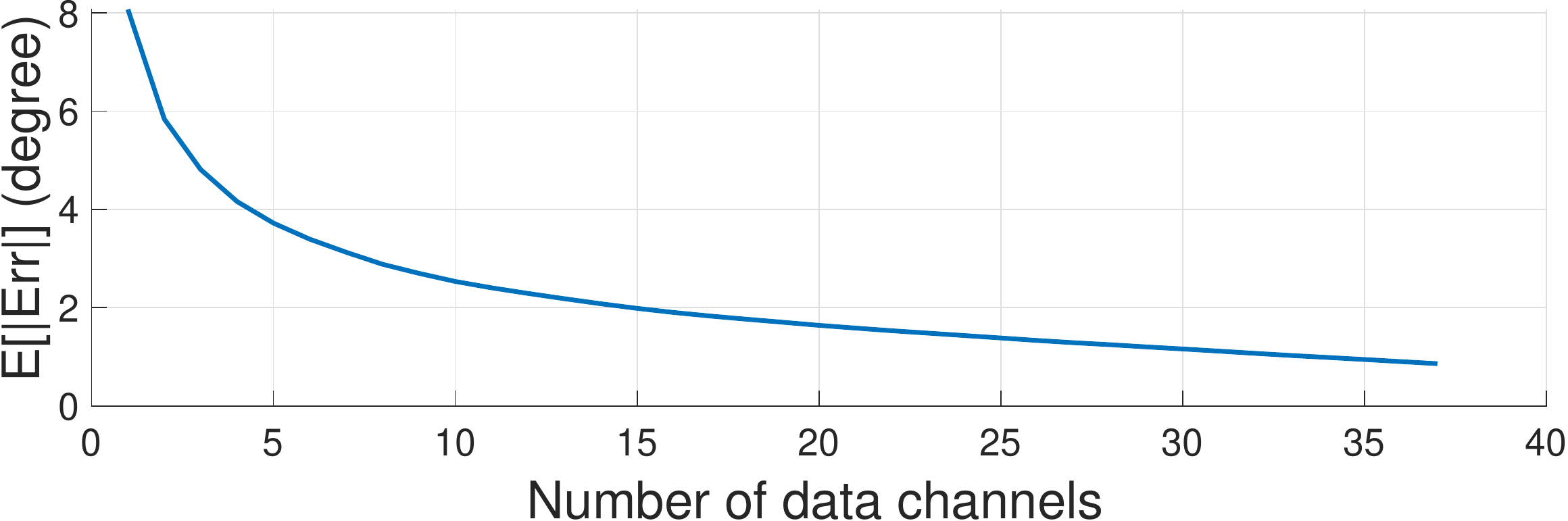}
\caption{The effect of number of channels used for AoA estimation on the average absolute error.}
\label{fig:fhmeanindoor}
\end{figure}

We complete our AoA analysis by examining the average absolute error that we may obtain with the same dataset in case we limit the average to a restricted number of data channels. For each number we compute the average error by considering all possible FH sequences with a different starting channel and different hop value. We depict the obtained results in Fig.~\ref{fig:fhmeanindoor}. We observe that starting with 15 channels, the average estimation error is well below $2\degree$.

\subsection{Positioning Accuracy}

To evaluate the accuracy of 2D positioning based on AoA detection, we conduct experiments using the playground area shown in Fig.~\ref{fig:playground}. We use the two receivers $A_1$ and $A_2$ placed in the bottom and left corners of the shaded area, within which we constrain the position of the transmitter according to the angular accuracy limitations identified and discussed in the previous subsection. We consider a four by five position grid spanning 4 m over the x- and 2.7 m over the y-axis. The system operates as before, by receiving 30 packets per channel and hopping over all data channels. 
After collecting the angular data generated by each receiver, we apply the methodology described in Sec.~\ref{sec:positioning} to compute positions $(x,y)$ of the target.

\begin{figure}
    \centering
    \includegraphics[width=0.99\columnwidth]{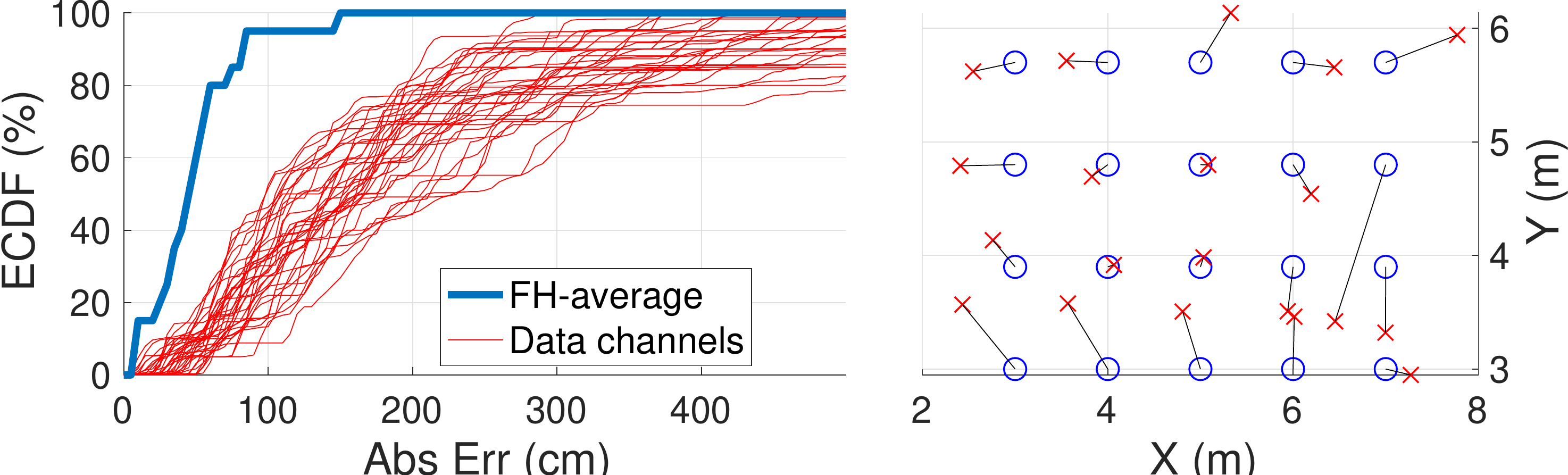}
    \caption{ECDF of indoor position estimation error (left). Qualitative estimation, real (circles) vs estimated (crosses) positions.}
    \label{fig:xyecdfandpositioning}
\end{figure}

We quantify the positioning accuracy in Fig.~\ref{fig:xyecdfandpositioning} (left), where we report the ECDF of the absolute estimation error. Observe that (thick blue line) the error is below 85~cm for more than 95\% of the positions. However, this is far from meeting the centimetre level accuracy expected by IoT applications, since the absolute positioning error is $<$10~cm only in 15\% of cases. We report on the right a qualitative evaluation: we show with blue circles and red crosses the real and estimated positions.

\section{Compromising AoA Estimation}

At this early {\it development stage}, the AoA mechanism in the BLE 5.1 standard does not enforce any security provisions. Surprisingly, no procedure for detecting whether interference affects the transmission of the CTE is considered. Indeed, this follows the CRC that protects the packet, thus there is no way of checking CTE correctness. This offers attackers opportunities to exploit the AoA based positioning capability for malicious purposes, as we explain next.

\begin{figure}[b]
    \centering
    \includegraphics[width=0.82\columnwidth]{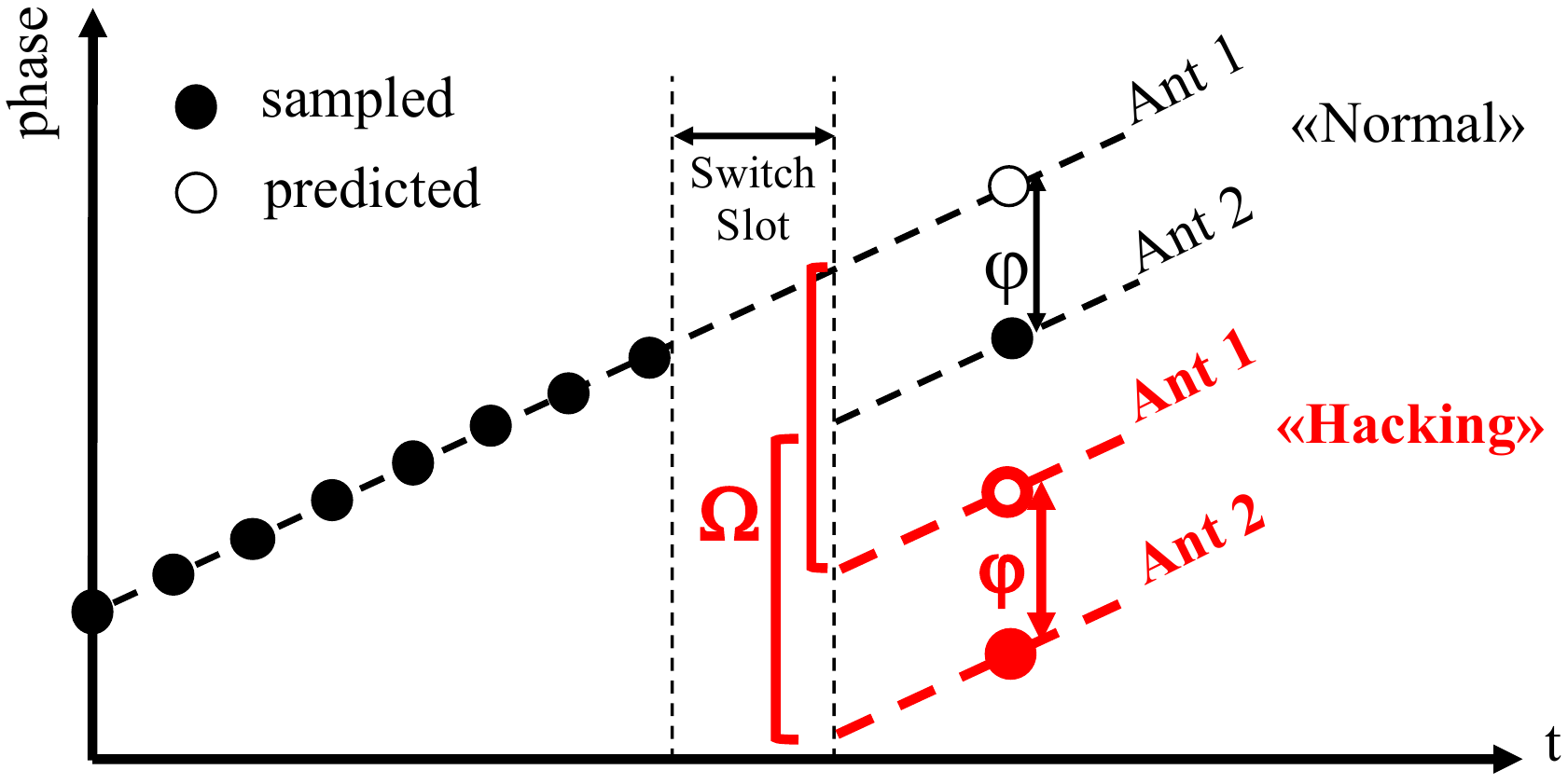}
    \caption{The transmitter can control the phase-delay computed at receiver by modifying phase after switching time, thus compromising correct AoA detection.}
    \label{fig:hacker}
\end{figure}

As we show in Fig.~\ref{fig:hacker}, in our implementation we compute a single value for $\varphi$ by subtracting the phase on antenna 2 (filled circle) from that predicted on antenna 1 (empty circle). Despite the very low level of complexity -- in a real device this technique requires just one integrated Single-Pole Double Throw circuit switch (SPDT) -- we showed in the previous section some good results that could attract manufacturers. However, we will demonstrate a simple attack on this procedure and propose simple countermeasures. To this end, we change the code of the transmitter to artificially modify the phase of the CTE during the switch slot: here we anticipate the phase of a constant term $\Omega$.

The value predicted by the AoA method on antenna 1 does not change, since the receiver keeps assuming its value corresponds to the top empty circle in Fig.~\ref{fig:hacker} (``Normal'' case). However, the value sampled on antenna 2 is different and corresponds to the bottom filled circle (``Hacking'' case). The computed phase-delay is hence $\varphi + \Omega$. This gives the transmitter the ability to modify the detected angle over time by properly choosing $\Omega$. 

\begin{figure}
    \centering
    \includegraphics[width=0.92\columnwidth]{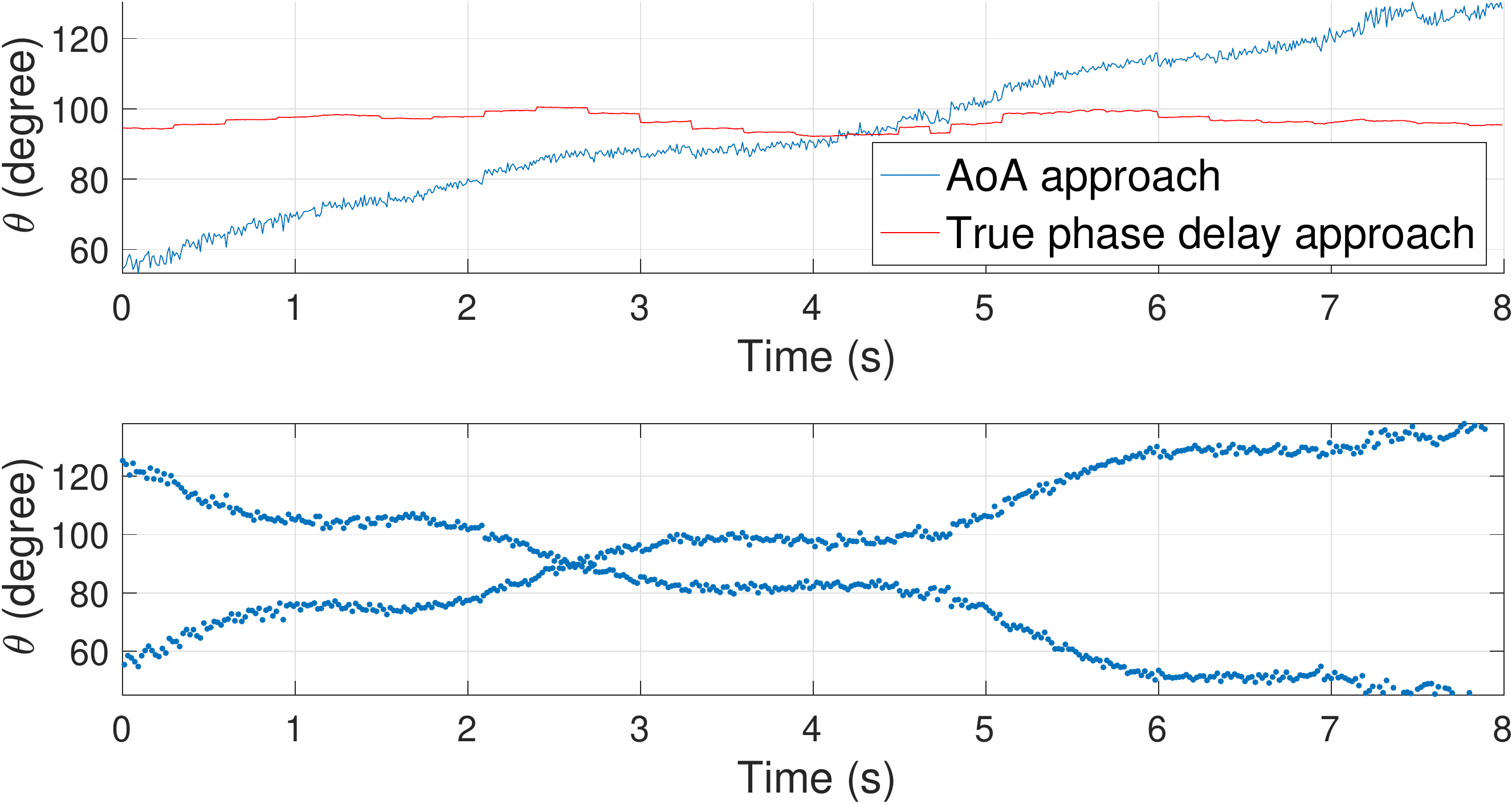}
    \caption{By artificially modifying the signal phase, a transmitter can trick a BLE 5.1 receiver into measuring an arbitrary crafted angle even if not moving (top). A simple approach to detect misleading devices~involves changing  antenna switching pattern (bottom).}
    \label{fig:fake_angle}
\end{figure}

To demonstrate the feasibility of this attack, we run an experiment with the transmitter placed in front of the AoA receiver sweeping $\Omega$ linearly over time, between $-\pi/6$ and $\pi/6$, which corresponds to a rotation of approximately $60\degree$ around the receiver. We report the detected angle $\theta$ with the blue line in Fig.~\ref{fig:fake_angle}-top. We want to underline that the same procedure can be adopted to trick more complex receivers using multiple antennas, by modifying the signal phase multiple times during the transmission of the CTE.
For comparison we also show in the figure with a red line the angle that would be measured by a {\it classic} approach that receives the signals at the two antennas with two coherent radio chains active at the same time. Since no prediction is involved in this case, the receiver would not be tricked by any artificial modification of the signal phase.

As a simple countermeasure, we can slightly change the behavior of the receiver, so that instead of using one main antenna and switching to the other only for measuring the phase-delay, it keeps the other antenna active for the next packet to be received. In this case, the resulting angle $\theta$ appears instead as in Fig.~\ref{fig:fake_angle}-bottom and an untruthful transmitter would be immediately discovered. Needless to say that should the transmitter know the switching pattern, it would always be able to properly craft the phase. To prevent this, keeping the switching pattern hidden to the transmitter and not deterministic would be an effective way of detecting such positioning attacks. 


\section{Summary \& Conclusions}\label{sec:conclusions}

In this paper, we performed an empirical evaluation of the AoA based positioning mechanism incorporated in the BLE 5.1 standard. We revealed that angular detection accuracy is limited to a constrained range and localization within few centimeters remains difficult. We further showed that an attacker may tamper with the BLE packet structure to mislead the positioning system, and we proposed simple guidelines that manufactures can implement to guarantee the truthfulness of this feature.


\bibliographystyle{unsrt} 

\end{document}